\documentclass[12pt,fleqn]{article}

\usepackage{latexsym}
\usepackage{amsmath}
\usepackage{amsthm}
\usepackage{amssymb}
\usepackage{amsfonts}

\begin{document}

\renewcommand{\]}{$$%
\par  \noindent  \hspace{-0.4em}}

\newcommand{\rf}[1]{(\ref{#1})}
\newcommand{\rff}[2]{(\ref{#1}\ref{#2})}

\newcommand{\ba}{\begin{array}}
\newcommand{\ea}{\end{array}}

\newcommand{\be}{\begin{equation}}
\newcommand{\ee}{\end{equation}}

\newcommand{\const}{{\rm const}}

\newcommand{\Cl}{{\cal C}}
\newcommand{\E}{\mathbb E}
\newcommand{\C}{\mathbb C}
\newcommand{\R}{\mathbb R}

\newcommand{\e}{{\bf e}}

\newcommand{\m}{\left( \ba{l}}
\newcommand{\ema}{\ea \right)}
\newcommand{\mm}{\left( \ba{cc}}
\newcommand{\miv}{\left( \ba{cccc}}

\newcommand{\scal}[2]{\mbox{$\langle #1 \! \mid #2 \rangle $}}

\newtheorem{prop}{Proposition}
\newtheorem{Th}{Theorem}  
\newtheorem{lem}{Lemma}
\newtheorem{rem}{Remark}
\newtheorem{cor}{Corollary}
\newtheorem{Def}{Definition}
\newtheorem{open}{Open problem}

\title{\bf 
A geometric interpretation of the spectral 
parameter for surfaces of constant mean curvature}

\author{
 {\bf Jan L.\ Cie\'sli\'nski}
\\ {\footnotesize Uniwersytet w Bia\l ymstoku,  
Instytut Fizyki Teoretycznej}
\\ {\footnotesize ul.\ Lipowa 41, 15-424  
Bia\l ystok, Poland}
\\ {\footnotesize e-mail: \tt janek\,@\,alpha.uwb.edu.pl }} 

\date{}

\maketitle

\begin{abstract}
Considering the kinematics of the moving frame associated with 
a constant mean curvature surface immersed in $S^3$ we derive 
a linear problem with the spectral parameter corresponding to 
elliptic sinh-Gordon equation. The spectral parameter is related 
to the radius $R$ of the sphere $S^3$. The application of 
the Sym formula to this linear problem yields
constant mean curvature surfaces in $\E^3$. Independently, we show that the Sym formula itself 
can be derived by an appropriate limiting process $R \rightarrow \infty$. 
\end{abstract}

\vspace{1cm}

\noindent {\it Mathematics Subject Classification 2000:} 
\par Primary 53A07, Secondary 37K25, 37K35.

\noindent {\it Key words and phrases:} \par
integrable systems, constant mean curvature surfaces, 
spectral parameter, Sym formula, elliptic sinh-Gordon equation

\pagebreak

Integrable nonlinear equations in $1+1$ dimensions are distinguished 
by the existence of the linear problem or spectral problem, 
i.e., an associated system of linear equations,
containing the so called spectral parameter 
(see, for instance, \cite{ZMNP}). 
The integrability conditions for the linear problem 
are equivalent to the considered nonlinear system. 
Integrable systems   played an important role in 
the classical differential geometry \cite{Eis1}, and 
are more and more important in the modern differential geometry 
\cite{Ch,Ten-book,TU}. Some integrable systems are of geometric origin \cite{Bob,FG,RS,Sym}.

Given a spectral problem we can construct 
a local immersion by the so called Sym formula 
\cite{Ci-FG,Sym}.
For instance, starting from the spectral problem for the 
sine-Gordon equation we get pseudospherical surfaces. 
The Sym approach gives probably the best correspondence 
between the geometry and spectral problems \cite{Ci-nos,Sym}. 
The spectral problem is necessary for 
the application of various methods of the soliton theory, like the inverse scattering method, 
the Darboux-B\"acklund transformation or algebro-geometric solutions in terms of 
Riemann theta functions. The Sym formula allows one 
to use all these methods in differential geometry. 

In the differential geometry of immersed submanifolds we have 
always a typical pair: 
the linear system of Gauss-Weingarten equations and their 
compatibility conditions, the nonlinear system of Gauss-Codazzi-Ricci equations. To obtain a linear problem of the 
soliton theory we need to insert a spectral parameter into the Gauss-Weingarten equations under consideration 
(for more details and references see, for instance, 
\cite{Ci-FG}). 

In this paper we consider surfaces of constant mean 
curvature $H \neq 0$. Constant mean curvature surfaces immersed in 3-dimensional Euclidean space $\E^3$ appear 
in the problem of soap bubbles if the (constant) outer pressure on 
both sides of the bubble surface is different 
(if the pressure is the same on both sides, we get minimal surfaces, i.e., $H = 0$).
The Gauss map for constant mean curvature surfaces in $\E^3$ 
is harmonic, i.e., 
the normal vector $N$ satisfies 
the equation
\be  \label{Nmodel}
     N,_{xx} + N,_{yy} + (N,_x^2 + N,_y^2) N = 0 \ , \qquad N^2 = 1 \ ,
\ee
where $x, y$ are curvature coordinates (or their conformal equivalents).
This is the 2-dimensional Euclidean $O(3)$ $\sigma$-model which appears in the classical field theory 
\cite{PZ} and also describes static solutions 
of $2+1$-dimensional continuum classical Heisenberg ferromagnet equation
\be
S,_t = S \times (S,_{xx} + S,_{yy}) \ , \qquad S^2 = 1 \ .
\ee

Here we use the approach 
proposed by Doliwa and Santini \cite{DS0} which has been   
successfully applied to the case of submanifolds of negative constant sectional curvature \cite{AC-vuz,CA}.
The Gauss-Weingarten equations 
for surfaces of constant mean curvature immersed 
in the sphere $S^3 \subset \E^4$ contain explicitly  the radius $R$ of the sphere $S^3$.
The main result of this paper is to show that $R$ plays the role of the spectral parameter (or, more precisely, $R$ is a function of the spectral parameter).

We consider an immersion in the sphere $S^3 \subset E^4$  
(of radius R) defined by the position 
vector $r = r(x,y)$. 
The unit vector $r/R$ is orthogonal to $S^3$ and 
we choose the second normal vector $n$ to be tangent to $S^3$.  
The immersion has 2-dimensional normal space
spanned by $r/R$ and $n$. 
We can always consider conformal coordinates, i.e., such that the
first fundamental form is proportional to $dx^2 + dy^2$, while the second fundamental form associated with $n$ is arbitrary:
\be  \ba{l}  \label{I,II}
I :=  dr\cdot dr = e^{2 \vartheta} ( dx^2 +  dy^2 ) \ , \\[2ex]
II := - dr\cdot dn = b_{11} dx^2 + 2 b_{12} dx dy + b_{22} dy^2 \ .
\ea \ee
Moreover, the second fundamental form $II'$ associated
with the normal $r/R$ is proportional to the metric. Indeed,
\be  \label{II'}
II' := - dr\cdot d(r/R) = - \frac{1}{R} dr\cdot dr = 
- k_0  e^{2 \vartheta} ( dx^2 +  dy^2 ) \ ,
\ee
where we denoted $k_0 := 1/R$. 

We denote unit tangent vectors by $E_1 \equiv e^{-\vartheta} r,_x$, 
$E_2 \equiv e^{-\vartheta} r,_x$, 
and the normal vectors by $E_3 \equiv n$, $E_4 \equiv r/R$.
The so called mean curvature vector (see, for instance, \cite{Am-subm,Chen-subm}) is given by
\be
    \vec H =  h E_3 - k_0 E_4 \ ,
\ee
where 
\be  \label{h}
 h := \frac{1}{2} \left(  b_{11} + b_{22} \right) e^{- 2\vartheta} \  .
\ee
We recall that in $d$-submanifold case ($d > 2$) the covariant constancy of $\vec H$ is a natural generalization of the condition $H = \const$.

Kinematics of the adapted frame (Gauss-Weingarten equations 
or structural equations) can be expressed in terms of the coefficients of the fundamental forms:
\be \ba{l} \label{GW}
r,_{xx} = \vartheta,_x  r,_x - \vartheta,_y  r,_y +  b_{11} n 
- R^{-2} e^{2 \vartheta} r \ , \\[2ex] 
r,_{xy} = \vartheta,_y  r,_x + \vartheta,_x  r,_y   +   b_{12} n \ , \\[2ex]
r,_{yy} = \vartheta,_y  r,_y - \vartheta,_x  r,_x 
+ b_{22} n - R^{-2} e^{2 \vartheta} r \ , \\[2ex] 
n,_x = - b_{11} e^{-2 \vartheta}  r,_x \ , \quad
n,_y = - b_{22} e^{-2 \vartheta}  r,_y \ .
\ea \ee
Note that $n \equiv E_3$ and $E_4 \equiv k_0 r$ are 
covariantly constant. Therefore  
$\vec H$ is covariantly constant iff $h = \const$.

We can rewrite the equations \rf{GW} in the matrix form:
\be \ba{l} \label{prlin}
\frac{\partial}{\partial x} \m E_1 \\ E_2 \\ E_3 \\ E_4 \ema =
\miv 0 & - \vartheta,_y  &  b_{11} e^{-\vartheta}  & -  k_0  e^{\vartheta}  \\
\vartheta,_y  & 0 & b_{12} e^{-\vartheta}   & 0 \\ - b_{11} e^{-\vartheta} 
  & - b_{12} e^{-\vartheta}   & 0 & 0 \\
k_0  e^{\vartheta}   & 0 & 0 & 0 \ema 
\m E_1 \\ E_2 \\ E_3 \\ E_4  \ema \ ,  \\[7ex]
\frac{\partial}{\partial y} \m E_1 \\ E_2 \\ E_3 \\ E_4  \ema =
\miv 0 & \vartheta,_x & b_{12} e^{-\vartheta}  & 0 \\
- \vartheta,_x  & 0 &  b_{22} e^{-\vartheta}   & - k_0  e^\vartheta \\ 
- b_{12} e^{-\vartheta}  & - b_{22} e^{-\vartheta} & 0 & 0 \\
0 & k_0 e^\vartheta  & 0 & 0 \ema 
\m E_1 \\ E_2 \\ E_3 \\ E_4  \ema \ .
\ea \ee
Denoting $(E_1, E_2, E_3, E_4)^T \in SO(4)$  by $\Phi$ and 
standard generators of the matrix Lie algebra $so(4)$ by $f_{jk}$, we have 
\be  \label{GW-so4}
\ba{l} \displaystyle
  \Phi,_x = \hat U \Phi \equiv 
\left( - \vartheta,_y  f_{12} + b_{11} e^{-\vartheta} f_{13} 
- k_0 e^\vartheta  f_{14} + b_{12} e^{-\vartheta} f_{23} \right) \Phi \ , \\[3ex] \displaystyle
\Phi,_y = \hat V \Phi \equiv 
\left( \vartheta,_x f_{12} + b_{22} e^{-\vartheta} f_{23} - 
k_0 e^\vartheta  f_{24} + b_{12} e^{-\vartheta} f_{13} \right) \Phi \ .
\ea \ee
The system of Gauss-Codazzi equations (identical with the compatibility
conditions for the above system of matrix linear equations)
is given by
\be \ba{l} 
\vartheta,_{xx} + \vartheta,_{yy} + (b_{11} b_{22} - b_{12}^2 ) e^{- 2 \vartheta} 
+ k_0^2 e^{2 \vartheta} = 0 \ , \\[2ex] 
b_{12},_x = b_{11},_y - \vartheta,_y ( b_{11} + b_{22} ) \ ,  \\[2ex]
b_{12},_y = b_{22},_x - \vartheta,_x ( b_{11} + b_{22} ) \ .
\ea \ee

Introducing  complex variables $z = x + i y$, $\bar z = x - i y$ and 
a complex function 
\be \label{Q}
  Q := \frac{1}{4} \left( b_{11} - b_{22} \right) - \frac{1}{2} i b_{12} \ ,
\ee 
known as the Hopf differential (compare \cite{Bob} where the case of 
surfaces in $\E^3$ is discussed in detail), we may rewrite the Gauss-Codazzi 
equations as
\be  \ba{l}  \label{GMCcompl}
4 \vartheta,_{z \bar z} +  (h^2 + k_0^2) e^{2\vartheta} 
- 4 Q {\bar Q} e^{-2\vartheta} = 0 \ ,  \\[2ex]
 Q,_{\bar z} = \frac{1}{2} h,_z e^{2 \vartheta} \ .
\ea \ee

If $h = \const$, then $r(x,y)$ describes constant mean curvature surfaces immersed in $\E^4$ and 
the system \rf{GMCcompl} reduces to 
\be  \label{GMC-H}
\vartheta,_{z \bar z} + \frac{1}{4} H^2  e^{2\vartheta} -  Q {\bar Q} e^{-2\vartheta} = 0 \ ,  \qquad
 Q = Q (z) \ ,
\ee
(i.e., $Q(z)$ is an analytic function) where
\be  \label{Pit}
   H^2 = h^2 + k_0^2 \ .
\ee 
The system \rf{GMC-H} can be transformed into 
the elliptic sinh-Gordon equation
\be  \label{sinh}
u,_{XX} + u,_{YY} = - \sinh u \cosh u 
\ee
by the change of variables $z \rightarrow Z \equiv X + i Y$ and $\theta \rightarrow u$, where
\be  \label{chanvar}
d Z = \sqrt{ 8 H Q(z) } \ dz \ , \qquad
u = \vartheta - \ln ( 2 H^{-1} |Q (z)| ) \ .
\ee
Note that the sign on the left hand side of \rf{sinh} 
is negative. Both cases, positive and negative, have some
applications in physics, and both are integrable 
(see, for instance, \cite{JK}).
The equations \rf{GMC-H} 
can also be interpreted as Gauss-Codazzi equations 
for surfaces of constant mean curvature $H$ immersed in 
$\E^3$. 

By virtue of \rf{Pit}, for any fixed $H$ linear equations 
\rf{GW-so4} 
form a one-parameter family of equations (the linear problem with the spectral parameter) parameterized by $\kappa$, 
where
\be  \label{hk0}
  h = H \cos\kappa \ , \qquad k_0 = H \sin\kappa \ ,
\ee
or, even better, $h$ and $k_0$ can be expressed 
in terms of $\zeta = e^{i\kappa}$, i.e.,
\be
 h = \frac{H}{2} \left( \zeta + \frac{1}{\zeta} \right) 
\ , \qquad \frac{1}{R} \equiv k_0 = 
\frac{H}{2 i} \left( \zeta - 
\frac{1}{\zeta} \right) \ .
\ee
We point out that the spectral parameter 
$\zeta$ takes values in the unit circle. 
The coefficients $b_{ij}$ appearing in equations
\rf{GW-so4}, can be expressed in terms of $Q$ and 
$\vartheta$ (compare \rf{h} and \rf{Q}):
\be   \label{bij}
b_{11} = h e^{2\vartheta} + 2 {\rm Re} Q \ , \quad
b_{12} = - 2 {\rm Im} Q \ , \quad 
b_{22} = h e^{2\vartheta} - 2 {\rm Re} Q \ .
\ee
Finally, we have the following $SO(4)$-valued spectral problem 
\be  \label{problin2} \displaystyle
\Phi,_x = {\hat U} \Phi \equiv \sum_{i<j} u_{ij} f_{ij} \Phi \ , 
\qquad
\Phi,_y = {\hat V} \Phi \equiv \sum_{i<j} v_{ij} f_{ij} \Phi \ , 
\ee
where
\be \ba{l}  \label{problin20} \displaystyle
\hat U = \frac{\zeta H}{2} e^\vartheta (f_{13} + i f_{14}) 
+ \frac{H}{2\zeta} e^\vartheta (f_{13} - i f_{14}) 
+ {\hat U}_0 \ , \\[3ex]
{\hat U}_0 := - 
\vartheta,_y f_{12} + 2 e^{-\vartheta} f_{13} {\rm Re} Q 
- 2 e^{-\vartheta} f_{23} {\rm Im} Q \ , \\[3ex]
\displaystyle
\hat V = \frac{\zeta H}{2} e^\vartheta (f_{23} + i f_{24}) 
+ \frac{H}{2\zeta} e^\vartheta (f_{23} - i f_{24}) 
+ {\hat V}_0 \ , \\[3ex]
{\hat V}_0 :=  
\vartheta,_x  f_{12} - 2 e^{-\vartheta} f_{23} {\rm Re} Q 
- 2 e^{-\vartheta} f_{13} {\rm Im} Q 
\ea \ee
The compatibility conditions for the linear problem 
\rf{problin2}, \rf{problin20} are given by \rf{GMC-H}. 
For our purposes it is very convenient to use the isomorphism 
$so(4) \simeq {\rm spin} (4)$, i.e., 
\be
   f_{ij} \qquad \longleftrightarrow \qquad 
  \frac{1}{2} \e_i \e_j \ ,
\ee
where $\e_j$ are so called Clifford numbers (see, for instance,\cite{Lou}), satisfying 
\be
 \e_j^2 = 1 \quad (j=1,2,3,4) \ , \qquad 
\e_i \e_j = - \e_j \e_i \qquad (i \neq j) \ .
\ee
The group ${\rm Spin} (4)$ is the double 
covering of $SO(4)$. 
Using the above isomorphism we immediately obtain the 
following ${\rm Spin}(4)$-valued spectral problem
\be \ba{l} \displaystyle  \label{problinspin}
\Psi,_x = U \Psi = 
\frac{1}{2} \sum_{i<j} u_{ij} \e_i \e_j \Psi \ , \\[3ex]  \displaystyle \Psi,_y = V \Psi = 
\frac{1}{2} \sum_{i<j} v_{ij} \e_i \e_j \Psi \ ,
\ea  \ee
where $u_{ij}$ and $v_{ij}$ are defined by \rf{problin2}, 
\rf{problin20}, i.e.,
\be \ba{l}  \label{problinspin1}
2 U = H e^\vartheta \e_1 (\e_3 \cos\kappa - \e_4 \sin\kappa) 
- \vartheta,_y \e_1 \e_2 + 2 e^{-\vartheta} (\e_1 {\rm Re} Q 
- \e_2 {\rm Im} Q ) \e_3 \ , \\[2ex] 
2 V = H e^\vartheta \e_2 (\e_3 \cos\kappa - \e_4 \sin\kappa) 
+ \vartheta,_x \e_1 \e_2 - 2 e^{-\vartheta} (\e_1 {\rm Im} Q 
+ \e_2 {\rm Re} Q ) \e_3 \ . \\[2ex] 
\ea \ee
One can easily check that the matrices of the spectral 
problem \rf{problinspin1} have the following 
 property:
\be  \label{invol}
 U (-\kappa) = \e_4 U (\kappa) \e_4^{-1} \ , \qquad
 V (-\kappa) = \e_4 V (\kappa) \e_4^{-1} \ .
\ee
Therefore we can confine ourselves to solutions $\Psi$ satisfying 
\be  \label{invol1}
 \Psi (-\kappa) = \e_4 \Psi (\kappa) \e_4^{-1}
\ee
Actually, $\kappa$ is defined as a positive quantity 
(compare \rf{hk0}). Therefore, the equations \rf{invol} 
can be treated as an extension of the obtained spectral problem 
on negative values of $\kappa$ (note that 
$\kappa < 0$ 
formally means that $R$ is negative as well).

The frame $E_k$ associated with the immersion $r$ can also be 
expressed in terms of Clifford numbers, namely
\be
  E_k  \quad \longleftrightarrow \quad {\bf E}_k := \Psi^{-1} \e_k \Psi \ .
\ee
Note that ${\bf E}_k$ form an orthonormal basis in the 
4-dimensional linear space $W$ spanned by $\e_1, \e_2, \e_3, \e_4$.
We define ${\bf \Phi} := ({\bf E}_1, {\bf E}_2, {\bf E}_3, {\bf E}_4)^T$ and compute 
\be
  {\bf E}_k,_x = (\Psi^{-1} \e_k \Psi),_x = 
  \Psi^{-1} [ \e_k, U ] \Psi \ ,
\ee
and similar expression for the $y$-derivative. By virtue of
\be
  [ \e_k, U] = 
\sum_{j > k}  u_{kj} \e_j  - \sum_{i < k}  u_{ik} \e_i \ ,
\ee
we get again the considered $SO(4)$ spectral problem (defined on the space $W \simeq {\mathbb E}^4$). Thus we proved 
that $\bf \Phi$ 
satisfies \rf{problin2}, \rf{problin20} provided that   
$\Psi$ solves \rf{problinspin} (compare an analogical  discussion for pseudospherical surfaces \cite{AC-vuz,CA}).

In particular, using $W \simeq {\mathbb E}^4$, we can express  by $\Psi$ the position vector ${\bf r} \in \E^4$ of the considered immersion of constant mean curvature 
(${\bf r} \simeq r$):
\be
      {\bf r} = R {\bf E}_4 \equiv R \Psi^{-1} \e_4 \Psi 
\ee
Finally, we define (compare \cite{CA}):
\be \label{F}
   F = \lim_{R \rightarrow \infty} ( {\bf r} - R \e_4 ) \ .
\ee
We expect that $F$ should be an immersion into ${\mathbb E}^3$
(for $R \rightarrow \infty$ the sphere $S^3$ locally becomes  
${\mathbb E}^3$). The subtraction of $R \e_4$ in the definition 
\rf{F} means that we choose a more convenient origin of 
the reference frame (a fixed point of the ``blowing'' sphere instead of the center of the sphere). This is the North pole
or (for $R < 0$) the South pole.

The limit $R \rightarrow \infty$ means that 
$\zeta \rightarrow 1$, $\kappa \rightarrow 0$ and 
$k_0 \rightarrow 0$. In this limit the matrices $U, V$ 
given by \rf{problinspin1} do not contain $\e_4$.
From \rf{invol1} it follows that
\be  \label{comrel}
 \Psi_0  \e_4 = \e_4 \Psi_0 \ , \qquad
\Psi'_0 \e_4 = - \e_4 \Psi'_0 \ ,
\ee
where $\Psi_0 := \Psi (x,y,0)$ (i.e., $\Psi$ evaluated at $\kappa = 0$) and the prime means differentiation with respect to $\kappa$. Thee limit \rf{F} can be 
computed as follows:
\be
 F = \lim_{R \rightarrow \infty} 
R (\Psi^{-1} \e_4 \Psi - \e_4) = \lim_{k_0 \rightarrow 0} 
\frac{\Psi^{-1} \e_4 \Psi - \e_4}{k_0} \ .
\ee
Applying L'Hospital's rule, and \rf{hk0} and 
\rf{comrel}, we 
get 
\be  \label{Sym}
  F = ( - \Psi^{-1} \Psi,_{k_0} \Psi^{-1} \e_4 \Psi + 
\Psi^{-1} \e_4 \Psi,_{k_0} )|_{k_0 = 0}  = 
2 H^{-1} \e_4 \Psi^{-1} \Psi,_\kappa |_{\kappa = 0} \ .
\ee
Thus we derived the Sym formula. The factor $\e_4$ 
turns out to be quite convenient because 
it assures that $F = F (x,y)$ describes an immersion in 
the space ${\mathbb E}^3$ spanned by $\e_1, \e_2, \e_3$.

We expect that the 
fundamental forms for the immersion $F$ can be obtained 
from the fundamental forms for $r$  in the limit $k_0 
\rightarrow 0$. Indeed, we compute
\be  \ba{l}
F,_x = 2 H^{-1} \e_4 \Psi_0^{-1} U,_\kappa (0) \Psi_0 = 
- \e_4 e^\vartheta \Psi_0^{-1} \e_1 \e_4 \Psi_0  = 
e^\vartheta {\tilde \e}_1 \ , \\[2ex]
F,_y = 2 H^{-1} \e_4 \Psi_0^{-1} V,_\kappa (0) \Psi_0 = 
- \e_4 e^\vartheta \Psi_0^{-1} \e_2 \e_4 \Psi_0  = 
e^\vartheta {\tilde \e}_2  \ , \\[2ex]
N = \Psi_0^{-1} \e_3 \Psi_0 = {\tilde \e}_3 \ ,  \\[2ex]
N,_x = \Psi_0^{-1}[\e_3, U (0)] \Psi_0 = 
  - H e^\vartheta {\tilde \e}_1 - 2 e^{-\vartheta} 
({\tilde \e}_1 {\rm Re} Q - {\tilde \e}_2 {\rm Im} Q  ) \ , \\[2ex]
N,_y = \Psi_0^{-1}[\e_3, V (0)] \Psi_0 = 
  - H e^\vartheta {\tilde \e}_2 + 2 e^{-\vartheta} 
({\tilde \e}_1 {\rm Im} Q + {\tilde \e}_2 {\rm Re} Q  ) \ ,
\ea \ee
where $\tilde \e_k := \Psi_0^{-1} \e_k \Psi_0$.  
Taking into account that ${\tilde \e}_1$, $\tilde \e_2$, 
$\tilde \e_3$ form an orthonormal frame in ${\mathbb E}^3$, 
we obtain the fundamental forms \rf{I,II}, \rf{bij}:
\be  \ba{l}  \label{I,II'}
I  = e^{2 \vartheta} ( dx^2 +  dy^2 ) \ , \\[2ex]
II = (h e^{2\vartheta} + 2 {\rm Re} Q ) dx^2 
 - 4 {\rm Im} Q \ dx dy + 
( h e^{2\vartheta} - 2 {\rm Re} Q ) dy^2 \ .
\ea \ee
Computing the mean curvature (according to the formula 
\rf{h}) we verify  that the obtained 
$F$ is a surface of constant mean 
curvature in ${\mathbb E}^3$. 

It is interesting that the presented geometric interpretation yields constant mean curvature surfaces
in ${\mathbb E}^3$ when the Sym formula is applied directly to  
${\rm Spin} (4)$ spectral problem. Usually one 
expects to get surfaces in ${\mathbb E}^3$ from some 
${\rm Spin} (3)$ (i.e., $SU(2)$) spectral 
problem. It turns out that such approach is possible 
in our case as well.
Namely, we can use the well known isomorphism
$so(4) \simeq so(3) \oplus so(3)$ . For instance, 
we can define
\be \ba{l}
  2 f_1 = \e_1 \e_2 - \e_3 \e_4 \ , \quad
 2 f_2 = \e_1 \e_3 + \e_2 \e_4 \ , \quad
2 f_3 = \e_1 \e_4 - \e_2 \e_3 \ , \\[2ex]
2 g_1 = \e_1 \e_2 + \e_3 \e_4 \ , \quad
 2 g_2 = \e_2 \e_4 - \e_1 \e_3 \ , \quad
2 g_3 = \e_1 \e_4 + \e_2 \e_3 \ .
\ea \ee 
One can easily check that
\be
[f_1, f_2] = f_3 \ , \quad [g_1, g_2] = g_3 \ , 
\quad [f_j, g_k] = 0 \ , 
\ee
etc., which means that $f_1, f_2, f_3$ and 
$g_1, g_2, g_3$ span two copies of commuting 
 Lie algebras $so(3)$.
The projector 
\be  \label{Pr}
 P := \frac{1}{2} \left( 1 + \e_1 \e_2 \e_3 \e_4
\right) 
\ee
projects $so(4)$ onto $so(3)$ spanned by $f_1, f_2, f_3$, while $I-P$ projects $so(4)$ on the second copy of $so(3)$.
Namely
\be \ba{l}
  P \e_1 \e_2 = f_1 \ , \quad 
P \e_1 \e_3 = f_2 \ , \quad 
P \e_1 \e_4 = f_3 \ , \\[2ex]
P \e_3 \e_4 =  - f_1 \ , \quad
P \e_2 \e_4 = f_2 \ , \quad 
P \e_2 \e_3 = - f_3 \ .
\ea \ee
Note that  $P f_k = f_k$ and $P g_k = 0$ for $k=1,2,3$.

Performing the projection \rf{Pr} we transform the spectral 
problem \rf{problinspin}, \rf{problinspin1} into 
\be \ba{l} \label{problin3}
2 U = H e^\vartheta  (f_2 \cos\kappa - f_3 \sin\kappa) 
- \vartheta,_y f_1 + 2 e^{-\vartheta} f_2 {\rm Re} Q 
+ 2 e^{-\vartheta} f_3 {\rm Im} Q \ , \\[2ex] 
2 V = H e^\vartheta (- f_3 \cos\kappa - f_2 \sin\kappa) 
+ \vartheta,_x  f_1 + 2 e^{-\vartheta} f_3 {\rm Re} Q 
- 2 e^{-\vartheta} f_2 {\rm Im} Q 
\ . 
\ea \ee
The projection $I-P$, applied to \rf{problinspin}, 
\rf{problinspin1}, yields 
\be \ba{l} \label{problin3'}
2 U = H e^\vartheta  (- g_2 \cos\kappa - g_3 \sin\kappa) 
- \vartheta,_y g_1 - 2 e^{-\vartheta} g_2 {\rm Re} Q 
- 2 e^{-\vartheta} g_3 {\rm Im} Q \ , \\[2ex] 
2 V = H e^\vartheta (g_3 \cos\kappa - g_2 \sin\kappa) 
+ \vartheta,_x  g_1 - 2 e^{-\vartheta} g_3 {\rm Re} Q 
+ 2 e^{-\vartheta} g_2 {\rm Im} Q 
\ . 
\ea \ee
The spectral problem \rf{problin3'} can be obtained 
from \rf{problin3} by a simple transformation, 
namely: $g_2 \rightarrow -f_2$, $g_3 \rightarrow -f_3$, 
$\kappa \rightarrow - \kappa$. 

We can apply the Sym formula 
$F = 2 H^{-1} \Psi^{-1} \Psi,_\kappa$ to the  
linear problems \rf{problin3} and \rf{problin3'}. 
Now we do not confine ourselves to 
$\kappa = 0$.  In both cases we get the same 
$\kappa$-family 
of surfaces of constant mean curvature $H$, 
characterized by the following fundamental forms:
\be  \ba{rcl}  \label{I,IIkap}
I & = & e^{2 \vartheta} (dx^2 + dy^2) \ , \\[2ex] 
II & = & ( H e^{2\vartheta} + 2(\cos\kappa\ {\rm Re} Q + \sin\kappa\ {\rm Im} Q )) dx^2  \\[2ex]
 & + & 4 (\sin\kappa\ 
{\rm Re} Q - \cos\kappa\ {\rm Im} Q ) dx dy \\[2ex]
& + &  
( H e^{2\vartheta} - 2(\cos\kappa\ {\rm Re} Q +
\sin\kappa\ {\rm Im} Q )) dy^2 
\ea \ee
The coefficient $2 H$ in the Sym formula, the same as 
in \rf{Sym}, is not very important, but is necessary 
to get the mean curvature $H$. Note that 
for $\kappa = 0$ we get the fundamental forms 
\rf{I,II'}.

Therefore, we derived  spectral problems 
\rf{problinspin1} and \rf{problin3},  
containing the 
spectral parameter, associated (by the Sym formula) 
with surfaces of constant mean 
curvature in $\E^3$. 
The $2\times 2$ spectral problems presented in the existing literature can be reduced to the spectral problem \rf{problin3} by some obvious gauge transformations and 
changes of variables (like \rf{chanvar}), 
compare \cite{Bob,Ci-nos,DS}. 

It turns out that the role of Clifford algebras and 
{\rm Spin} groups is quite important in the derivation of the Sym formula (compare the above results with 
\cite{CA}). It would be very interesting to 
extend our approach on the submanifolds asociated with spectral problems defined in terms of Clifford algebras \cite{Ci-Spin}.

\end{document}